\documentstyle[times,psfig,pramana,floats]{ias}

\def\ni{\noindent}

\newcommand{\beb}{\begin{thebibliography}{999}}
\newcommand{\eeb}{\end{thebibliography}}
\newcommand{\bei}{\begin{itemize}}
\newcommand{\eei}{\end{itemize}}
\newcommand{\bef}{\begin{figure}}
\newcommand{\eef}{\end{figure}}
\newcommand{\ben}{\begin{enumerate}}
\newcommand{\een}{\end{enumerate}}
\newcommand{\beq}{\begin{equation}}
\newcommand{\eeq}{\end{equation}}
\newcommand{\ber}{\begin{eqnarray}}
\newcommand{\eer}{\end{eqnarray}}
\newcommand{\twidle}{\widetilde}

\begin{document}

\mark{{Neutrino Propagation}{Konar \& Das}}
\title{Neutrino Propagation in a Weakly Magnetized Medium}

\author{Sushan Konar}
\address{Department of Physics \& Meteorology, IIT, Kharagpur 721302, India}

\author{Subinoy Das}
\address{Department of Physics, Columbia University, New York 10027, U.S.A.}


\abstract{
Neutrino-photon processes, forbidden in vacuum, can take place in the presence 
of a thermal medium and/or an external electro-magnetic field, mediated by the
corresponding charged leptons (real or virtual). Such interactions affect the
propagation of neutrinos through a magnetized plasma. We investigate the neutrino-photon
absorptive processes, at the one-loop level, for massless neutrinos in a weakly
magnetized plasma. We find that there is no correction to the absorptive part of the 
axial-vector--vector amplitude due to the presence of a magnetic field to the linear 
order in the field strength. 
}
\maketitle

\section{Introduction}{\label{intro}}
 
Neutrino-Photon reactions, forbidden (or highly suppressed) in vacuum, for example 
the plasmon decay ($\gamma \rightarrow \nu \nu$) or the Cerenkov process 
($\nu \rightarrow \nu \gamma$) and the cross-processes, can become important in 
regions with very dense plasma and/or large-scale external magnetic fields such as 
encountered in the cosmological or the astrophysical context~\cite{raff}. In the 
standard model, these $\nu-\gamma$ processes, appearing at the one-loop level, do 
not occur in vacuum because they are kinematically forbidden and also because the 
neutrinos do not couple to the photons at the tree-level. In the presence of a 
medium or a magnetic field, it is the charged particles (real plasma particles or 
virtual particles excited by an external field) running in the loop which, when 
integrated out, confer their electro-magnetic properties to the 
neutrino~\cite{hari,oliv,ioan}. 
These processes also become kinematically allowed since the photon dispersion 
relation is modified in the presence of a medium and/or an external magnetic field 
opening up the phase space for the such reactions to take 
place~\cite{oliv,hard,orae,shai,gsh1,gsh2}. A thermal medium or/and an external 
magnetic field, thus, fulfill the dual purpose of inducing an effective neutrino-photon 
vertex and of modifying the photon dispersion relation (see Ref.~\cite{ioan} and references 
therein for a detailed review).  \\

\ni
The enhancement of $\nu-\gamma$ interactions by magnetic fields in an effective 
Lagrangian framework has been discussed in Ref.~\cite{shame}. And, it has also 
been shown that the $\nu-\gamma$ interaction in presence of a thermal medium 
induces a small effective charge to the neutrino and that the neutrino electro-magnetic 
vertex is related to the photon self-energy in the medium~\cite{nie3,alth}. Recently, 
this effective charge has been calculated, considering not only a thermal medium but 
also an external magnetic field, for neutrinos coupled to dynamical photons having 
$q_0=0$ and $|\vec{q}| \to 0$ \cite{qft3}. In the weak field limit this 
effective charge acquired by a neutrino is, in fact, proportional to the field 
strength and also depends on the {\it direction} of the neutrino propagation 
with respect to the direction of the magnetic field.  \\

\ni
Evidently, these processes modify the neutrino propagation through a magnetized
medium. In order to accurately estimate the neutrino fluxes coming from regions 
pervaded by dense plasma and strong magnetic fields (various astrophysical objects, 
for example) it is extremely important to study, in particular, the absorptive 
processes. In view of this, we study the absorptive part of the 1-loop polarization 
tensor of the $\nu-\gamma$ interaction, in the present work. It should be noted here 
that for most astrophysical systems, where the $\nu-\gamma$ processes acquire 
importance by virtue of large plasma density or the presence of 
magnetic fields (supernovae, newly born neutron stars or late stages of stellar 
evolution) the field strength is almost always smaller than the QED critical field. 
Therefore, the weak field limit ($e{\cal B} < m_e^2$ i.e, ${\cal B} \le 10^{13}$~Gauss) 
is appropriate for most astrophysical situations. \\

\ni
We find that the absorptive part {\it vanishes to linear order in ${\cal B}$}. This 
is quite significant because it contradicts the naive expectation and implies that
the first non-trivial correction would come from terms containing higher powers of
the field strength. More importantly, it signifies that the absorptive processes do 
not acquire any spatial anisotropy due to the presence of a weak magnetic field. 
This is quite interesting. Because the magnitude of the effective charge of the 
neutrinos (coming from the real-part of the one-loop amplitudes) is dependent on the 
relative direction of the neutrino momentum with respect to the magnetic field for 
any non-zero field strength. But no such effect is seen for the absorptive part. \\
  
\ni
The organization of the paper is as follows. In section-\ref{form} we discuss the 
basics of neutrino-photon effective action and the fermion propagators in a 
magnetized medium. Section-\ref{calc} contains the details of the calculation of 
the 1-loop diagram and in section-\ref{weak} we consider the weak-field limit.
Finally, we conclude with a discussion on the possible implications of our result 
in section-\ref{concl}

\section{Formalism}{\label{form}}

The off-shell electro-magnetic vertex function $\Gamma_{\mu}$ is defined in such a 
way that, for on-shell neutrinos, the $\nu \nu \gamma$ amplitude is given by:
\beq
{\cal M} = - i \bar{u}(k') \Gamma_{\mu} u(k) A^{\mu}(q),
\eeq
where, $q,k,k'$ are the momentum carried by the photon and the neutrinos respectively 
and $q=k-k'$. Here, $u(k)$ is the neutrino wave-function and $A^{\mu}$ stands for the 
electro-magnetic vector potential. In general, $\Gamma_{\mu}$ would depend on $k$, $q$, 
the characteristics of the medium and the external electro-magnetic field.  We shall, 
in this work, consider neutrino momenta that are small compared to the masses of the 
W and Z bosons allowing us to neglect the momentum dependence in the W and Z propagators. 
This is equivalent to lowest-order $G_F$ calculations and is justified for low-energy 
neutrinos and low temperatures and weak fields compared to the Fermi scale. Since, in 
this work we focus our attention to the possible astrophysical applications (notably 
in the context of the supernovae, the neutron stars or the late stages of stellar 
evolution) the characteristic temperatures or magnetic fields defined by the densities 
in such systems are much larger than those actually observed in reality. Therefore, 
for the low energy neutrinos the four-fermion interaction is given by the following 
effective Lagrangian:
\beq
{\cal L}_{\rm eff} 
= \frac{1}{\sqrt{2}} G_F 
  {\overline \nu} \gamma^{\mu} (1 - \gamma_5) \nu {\overline l_\nu}
  \gamma_{\mu} (g_{\rm V} - g_{\rm A} \gamma_5) l_\nu \,,
\eeq
where, $\nu$ and $l_\nu$ are the neutrino and the corresponding lepton field 
respectively. For electron neutrinos,
\ber
g_{\rm V} &=& 1 - (1 - 4 \sin^2 \theta_{\rm W})/2, \\
g_{\rm A} &=& 1 - 1/2;
\eer
where the first terms in $g_{\rm V}$ and $g_{\rm A}$ are the contributions from the 
W exchange diagram and the second one from the Z exchange diagram. Then the amplitude 
effectively reduces to that of a purely photonic case with one of the photons replaced 
by the neutrino current, as seen in the diagram in fig.~\ref{f:cher}. Therefore, 
$\Gamma_{\nu}$ is given by:
\beq
\Gamma_{\nu} 
= - \frac{1}{\sqrt{2}e} G_F \gamma^{\mu} (1 - \gamma_5) \,(g_{\rm V} \Pi_{\mu \nu} 
  - g_{\rm A} \Pi_{\mu \nu}^5) \,,
\eeq
where, $\Pi_{\mu \nu}^5$ represents the axial-vector--vector coupling and $\Pi_{\mu \nu}$ 
is the polarization tensor arising from the diagram in fig.~\ref{f:polr}. We have analyzed 
the Lorentz tensor structure of $\Pi_{\mu \nu}$, taking into account all the available 
symmetry, in an earlier work \cite{qft1}. Subsequently, a similar analysis has been 
carried out in \cite{sahu} too. Since the tensorial structure of $\Pi_{\mu \nu}^5$ would 
be similar in form to that of $\Pi_{\mu \nu}$, we do not repeat the analysis here. It 
should be mentioned that the tensorial structure of $\Pi^5_{\mu \nu}$, in this context, 
has been discussed in detail in \cite{schu,shai}.  \\

\ni
Because of the electro-magnetic current conservation, for the polarization tensor, we 
have the following gauge invariance condition:
\beq
q^{\mu} \Pi_{\mu \nu} = 0 = \Pi_{\mu \nu} q^{\nu}.
\eeq
\bef
\begin{center}{\mbox{\psfig{file=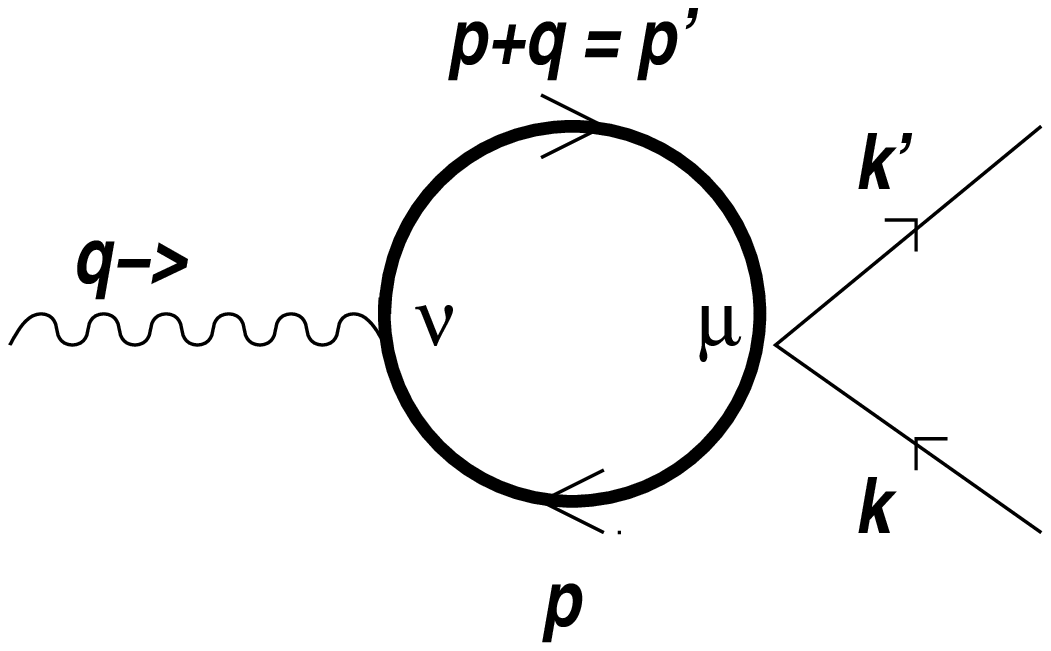,angle=-90,width=250pt}}}\end{center}
\caption{One-loop diagram for the effective electro-magnetic vertex of the neutrino 
in the limit of infinitely heavy W and Z masses.}{\label{f:cher}}
\eef
Same is true for the photon vertex of fig.~\ref{f:cher} and we have
\beq
\Pi_{\mu \nu}^5 q^{\nu} = 0\,.
\label{gi_pi5}
\eeq

\ni In an earlier paper \cite{qft2} we have calculated the imaginary part of 
$\Pi^{\mu \nu}$ in a background medium in presence of a uniform external magnetic 
field, in the weak-field limit, calculated at the 1-loop level. We shall use the 
results of \cite{qft2} here to obtain an expression for the total imaginary 
part of the effective neutrino current under equivalent conditions.  \\

\ni In order to calculate the absorptive processes in a thermal medium we use the 
real time formalism of the finite temperature field theory. The propagator acquires 
a matrix structure in this formalism and the off-diagonal elements provide the 
decay/production amplitudes. For the ease of calculation, we work with the 11-component 
of the propagator to find the imaginary part of the 11-component of the photon 
polarization tensor ($\Pi^{11}_{\mu \nu}$). This quantity, multiplied by appropriate 
factors, then gives the correct value of the imaginary part of the polarization 
tensor~\cite{adas,kob1,nie1,land}. Though for notational brevity we shall suppress 
the 11-superscript for both the propagator and the polarization tensor in the rest of 
the paper. It should be mentioned here that we consider the effective photon-neutrino 
interaction coming from the imaginary part of the axial-vector--vector amplitude. Hence, 
like in the case of the polarization tensor, we work with the imaginary part of the 
11-component of the axial-vector--vector amplitude.  \\

\bef
\begin{center}{\mbox{\psfig{file=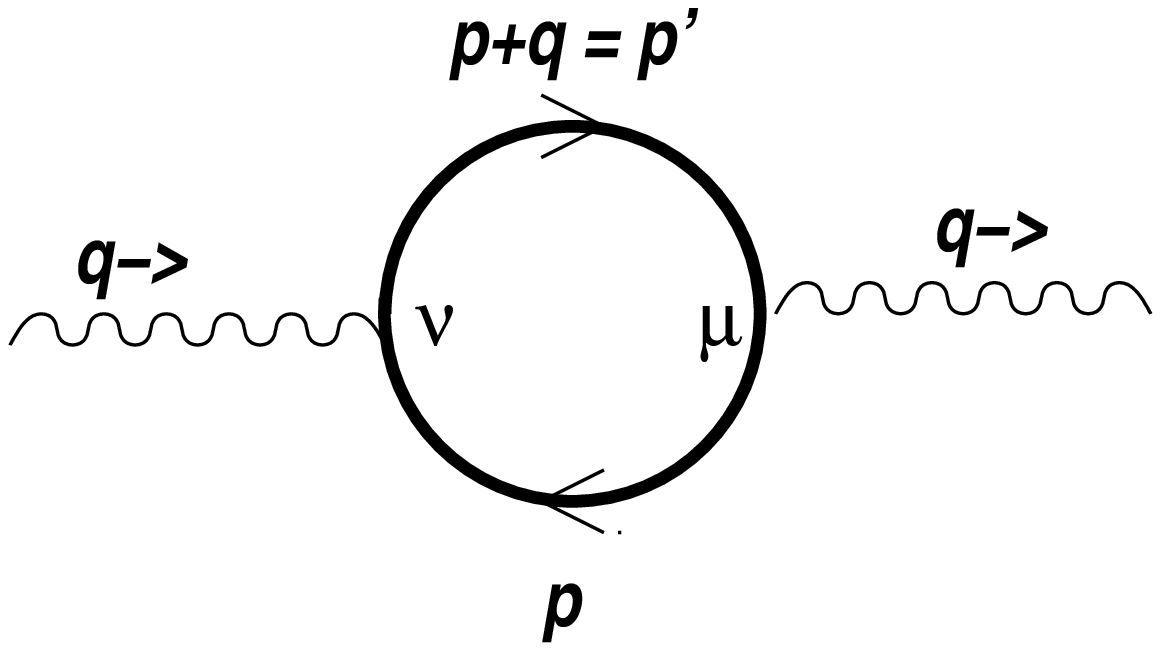,angle=-90,width=250pt}}}\end{center}
\caption[]{One-loop diagram for the vacuum polarization.}{\label{f:polr}}
\eef

\ni The dominant contribution to $\Pi_{\mu \nu}$ and $\Pi^5_{\mu \nu}$ come from the 
electron lines in the loop. To evaluate this diagram we use the electron propagator 
within a thermal medium in presence of a background electro-magnetic field. Rather 
than working with a completely general background field we specialize to the case 
of a constant (or slowly varying) magnetic field. Once this is assumed, the field can 
be taken in the $z$-direction without any further loss of generality. We denote the 
magnitude of this field by $\cal B$. Ignoring at first the presence of the medium, 
the electron propagator in such a field can be written down following Schwinger's 
approach~\cite{schw,tsai,ditt}:
\ber
i S_B^V(p) = \int_0^\infty ds\; e^{\Phi(p,s)} C(p,s) \,,
\label{SV}
\eer
where we have used the shorthands,
\ber
\Phi(p,s) &\equiv& is \left( p_\parallel^2 - {\tan (e{\cal B}s) \over e{\cal B}s} 
\, p_\perp^2 - m^2 \right) - \epsilon |s| \,, 
\label{Phi} \\
C(p,s) &\equiv& \Big[ ( 1 + i\sigma_z \tan  e{\cal B}s ) (\rlap/p_\parallel + m ) 
       - (\sec^2 e{\cal B}s) \rlap/ p_\perp \Big] \,. 
\label{C}
\eer
and
\ber
\rlap/ p_\parallel &=& \gamma_0 p_0 - \gamma_3 p_3 \\
\rlap/p_\perp &=& \gamma_1 p_1 + \gamma_2 p_2 \\
p_\parallel^2 &=& p_0^2 - p_3^2 \\
p_\perp^2 &=& p_1^2 + p_2^2 \,.
\eer
Also, $\sigma_z$ is given by:
\beq
\sigma_z = i\gamma_1 \gamma_2 = - \gamma_0 \gamma_3 \gamma_5 \,,
\label{sigz} 
\eeq
where the two forms are equivalent because of the definition of $\gamma_5$.
Of course in the range of integration indicated in Eq.(\ref{SV}) $s$ is never 
negative and hence $|s|$ equals $s$. It should be mentioned here that we follow 
the notation adopted in our previous papers~\cite{qft2,qft3} to ensure continuity. 
In the presence of a background medium, the above propagator is modified to~\cite{elmf}:
\beq
iS(p) = iS_B^V(p) - \eta_F(p) \left[ iS_B^V(p) - i\overline S_B^V(p) \right] \,,
\label{fullprop}
\eeq
where 
\beq
\overline S_B^V(p) \equiv \gamma_0 S^{V \dagger}_B(p) \gamma_0 \,,
\label{Sbar}
\eeq
for a fermion propagator and $\eta_F(p)$ contains the distribution function for the 
fermions and the anti-fermions:
\ber
\eta_F(p) = \Theta(p\cdot u) f_F(p,\mu,\beta) 
           + \, \Theta(-p\cdot u) f_F(-p,-\mu,\beta) \,.  
\label{eta} 
\eer
Here, $f_F$ denotes the Fermi-Dirac distribution function:
\beq
f_F(p,\mu,\beta) = {1\over e^{\beta(p\cdot u - \mu)} + 1} \,,
\eeq
and $\Theta$ is the step function. Rewriting Eq.(\ref{fullprop}) in the form:
\ber
iS(p) &=& iS_{\rm re} + iS_{\rm im}
\label{S_reim}
\eer
we recognize:
\ber
S_{\rm re} &=&  \frac{1}{2} \left[ S_B^V(p) + \overline S_B^V(p) \right] \, ,\\
S_{\rm im} &=& (1/2 - \eta_F(p)) \left[ S_B^V(p) - \overline S_B^V(p) \right] \, ;
\eer
where the subscripts {\em re} and {\em im} refer to the real and imaginary parts 
of the propagator. Using the form of $S_B^V(p)$ in Eq.(\ref{SV}) we obtain the 
imaginary part to be:
\ber
iS_{\rm im} &=& (1/2 - \eta_F(p)) \int_{-\infty}^\infty ds\; e^{\Phi(p,s)} C(p,s) \,.
\label{S_im}
\eer
with $\Phi(p,s)$ and $C(p,s)$ defined by Eq.s.(\ref{Phi}) and (\ref{C}).

\section{Calculation of the 1-loop Diagram}
\label{calc}

\ni {\bf $\Pi_{\mu \nu}(q, {\cal B})$ in odd powers of ${\cal B}$} - The amplitude 
of the 1-loop diagram of fig.~\ref{f:polr} can be written as:
\beq
i \Pi_{\mu\nu}(q) = - \int \frac{d^4p}{(2\pi)^4} (ie)^2 \; \mbox{tr}\, 
\left[\gamma_\mu \, iS(p) \gamma_\nu \, iS(p')\right] \,,
\eeq
where, for the sake of notational simplicity, we have used
\beq
p' = p+q \,.
\label{p'}
\eeq
The minus sign on the right side is for a closed fermion loop and $S(p)$ is the 
propagator given by Eq.(\ref{fullprop}). This implies that the absorptive part 
of the polarization tensor is given by:
\beq
\Pi_{\mu\nu}^{11}(q) 
= -ie^2 \int \frac{d^4p}{(2\pi)^4} \; \mbox{tr}\, 
  \left[\gamma_\mu \, iS_{\rm im}(p) \gamma_\nu \, iS_{\rm im}(p^{\prime}) \right] .
\label{PiT}
\eeq
And, the gauge invariant contribution to the absorptive part of the vacuum polarization 
tensor which is odd in $\cal B$ is given by \cite{qft2}:
\ber
{\Pi_{\mu\nu}(q, \beta)}^{\rm O}
&=& - \, 4ie^2 \varepsilon_{\mu\nu\alpha_\parallel\beta} q^\beta \, 
    \int \frac{d^4p}{(2\pi)^4} \, X(\beta, q, p) 
    \int_{-\infty}^\infty ds \; e^{\Phi(p,s)} \int_{-\infty}^\infty ds' \; e^{\Phi(p',s')} \nonumber\\*
&\times& \Bigg[ p^{\widetilde\alpha_\parallel} \tan e{\cal B}s 
    + p'^{\widetilde\alpha_\parallel} \tan e{\cal B}s' 
    - {\tan e{\cal B}s \; \tan e{\cal B}s' \over \tan e{\cal B}(s+s')} 
       \; (p+p')^{\widetilde\alpha_\parallel} \Bigg] \,,
\label{oddpart}
\eer
where we have defined:
\beq
X(\beta, q, p) = (1/2 - \eta_F(p)) \, (1/2 - \eta_F(p^{\prime})) \,.
\label{X}
\eeq
\ni {\bf $\Pi^5_{\mu \nu}(k, {\cal B})$ in odd powers of ${\cal B}$} - The amplitude 
of the 1-loop diagram of fig.~\ref{f:cher} can be written as:
\beq
\Pi^5_{\mu\nu}(q)
= -ie^2 \int \frac{d^4p}{(2\pi)^4} \; 
   \mbox{tr}\, \left[\gamma_\mu \, \gamma_5 \, iS(p) \gamma_\nu \, iS(p')\right] \,.
\label{1loopampl}
\eeq
Using Eq.(\ref{S_im}) we find that the absorptive part of the polarization tensor is given by:
\ber
\Pi^5_{\mu\nu}(q)
&=& -ie^2 \int \frac{d^4p}{(2\pi)^4} \; X(\beta, q, p) 
    \, \int_{-\infty}^\infty ds\; e^{\Phi(p,s)} \, 
    \int_{-\infty}^\infty ds^{\prime}\; e^{\Phi(p^{\prime},s^{\prime})} \nonumber \\
&&  \times \, \mbox{tr}\, \left[\gamma_\mu \, \gamma_5 \, C(p,s) \gamma_\nu \, 
    C(p^{\prime},s^{\prime}) \right] \,.
\label{Pi5}
\eer
Notice that the phase factors appearing in Eq.~(\ref{Pi5}) are even in $\cal B$. 
Thus, we need consider only the odd terms from the traces. Performing the traces, 
the expression, odd in powers of ${\cal B}$, comes out to be:
\ber
{\Pi^5_{\mu\nu}(q)}^{\rm O} 
= - 4e^2 \int \frac{d^4p}{(2\pi)^4} \; X(\beta, q, p) 
  \int_{-\infty}^\infty ds\; e^{\Phi(p,s)} \,
    \int_{-\infty}^\infty ds^{\prime}\; e^{\Phi(p^{\prime},s^{\prime})} \, R_{\mu \nu} \,,
\eer
where,
\ber
R_{\mu \nu} 
&=& \varepsilon_{\mu \nu 1 2} \, m^2 \, 
    (\tan  e{\cal B}s + \tan e{\cal B}s^{\prime}) \nonumber \\
&+& \left(g_{\mu \twidle{\alpha_\parallel}} \, g_{\nu \beta_\parallel} 
    + g_{\mu \beta_\parallel} \, g_{\nu \twidle{\alpha_\parallel}}\right) \,
    p^{\alpha_\parallel} \, p^{\beta_\parallel} 
    \, (\tan  e{\cal B}s + \tan  e{\cal B}s^{\prime}) \nonumber \\
&-& \left(g_{\mu \twidle{\alpha_\parallel}} \, g_{\nu \beta_\perp} \,
    + g_{\mu \beta_\perp} \, g_{\nu \twidle{\alpha_\parallel}}\right) 
    p^{\alpha_\parallel} \,  p^{\prime \beta_\perp} 
    \, \tan  e{\cal B}s \, \sec^2 e{\cal B}s^{\prime} \nonumber \\
&-& \left(g_{\mu \twidle{\alpha_\parallel}} \, g_{\nu \beta_\perp} \,
    + g_{\mu \beta_\perp} \, g_{\nu \twidle{\alpha_\parallel}}\right) 
    p^{\beta_\perp} \, p^{\prime \alpha_\parallel} 
    \, \tan  e{\cal B}s^{\prime} \, \sec^2 e{\cal B}s \nonumber \\
&+& \left(g_{\mu \twidle{\alpha_\parallel}} \, g_{\nu \beta_\parallel} 
    + g_{\mu \beta_\parallel} \, g_{\nu \twidle{\alpha_\parallel}}  
    - g_{\mu \nu} g_{\twidle{\alpha_\parallel} \beta_\parallel}\right) 
    \left(p^{\alpha_\parallel} \, q^{\beta_\parallel} \, \tan  e{\cal B}s
    + q^{\alpha_\parallel} \, p^{\beta_\parallel} \, \tan  e{\cal B}s^{\prime}\right) \,.
\label{R1}
\eer
In writing this expression, we have used the notation $g_{\mu \widetilde\alpha_\parallel}$, 
for example. This signifies that $\widetilde\alpha_\parallel$ is an index which can 
take only the `parallel' indices, i.e., 0 and 3, and is moreover different from the 
index $\alpha$ appearing elsewhere in the expression. Now, since we perform the 
calculations in the rest frame of the medium where $p\cdot u=p_0$ the distribution 
function does not depend on the spatial components of $p$. In the last two terms of 
Eq.(\ref{R1}), the integral over the transverse components of $p$ has the following 
generic structure:
\beq
\int d^2 p_\perp \; e^{\Phi(p,s)} e^{\Phi(p',s')} \times
\mbox{($p^{\beta_\perp}$ or $p'^{\beta_\perp}$)} \,.
\eeq
Notice that,
\ber
{\partial \over \partial p_{\beta_\perp}} \Big[ \; e^{\Phi(p,s)} e^{\Phi(p',s')} \Big]
= \Big( \tan e{\cal B}s \; p^{\beta_\perp} + \tan e{\cal B}s' \; p'^{\beta_\perp} \Big) 
  {2i\over e{\cal B}} e^{\Phi(p,s)} e^{\Phi(p',s')} \,.
\label{single_derivative}
\eer
However, this expression, being a total derivative, should integrate to zero. Thus we 
obtain that,
\beq
\tan e{\cal B}s \; p^{\beta_\perp} \stackrel\circ= - \tan e{\cal B}s' \; p'^{\beta_\perp} \,,
\eeq
where the sign `$\stackrel\circ=$' means that the expressions on both sides of it, 
though not necessarily equal algebraically, yield the same integral. This gives,
\ber
p^{\beta_\perp} &\stackrel\circ=& - \, {\tan e{\cal B}s' \over \tan e{\cal B}s
+ \tan e{\cal B}s'} \; q^{\beta_\perp} \,,\nonumber\\*
p'^{\beta_\perp} &\stackrel\circ=&  {\tan e{\cal B}s \over \tan e{\cal B}s
+ \tan e{\cal B}s'} \; q^{\beta_\perp} \,.
\eer
Also, using the definition of the exponential factor $\Phi(p,s)$ from Eq.(\ref{Phi}), 
we notice that
\ber
&& m^2 \tan e{\cal B}s \; e^{\Phi(p,s)} \, e^{\Phi(p',s')} \nonumber \\
&& = \tan e{\cal B}s \, \Big\{i {d \over ds'} + (p_\parallel^{\prime2} - \sec^2 e{\cal B}s' \; 
     p_\perp^{\prime2}) \Big\} \, e^{\Phi(p,s)} \, e^{\Phi(p',s')} \,.
\label{msq}
\eer
Moreover, taking another derivative with respect to $p^{\alpha_\perp}$ of 
Eq.(\ref{single_derivative}) we obtain, from the fact that this derivative 
should also vanish on $p$ integration, 
\ber
p_\perp^2 
\stackrel\circ= {1\over \tan e{\cal B}s + \tan e{\cal B}s'}
\Bigg[ -ie{\cal B} 
+ {\tan^2 e{\cal B}s' \over \tan e{\cal B}s + \tan e{\cal B}s'} \; q_\perp^2 \Bigg] \,.
\label{psq}
\eer
and,
\ber
{p'_\perp}^2 
\stackrel\circ= {1\over \tan e{\cal B}s + \tan e{\cal B}s'}
\Bigg[ -ie{\cal B} 
  + {\tan^2 e{\cal B}s \over \tan e{\cal B}s + \tan e{\cal B}s'} \; q_\perp^2 \Bigg] \,.
\label{p'sq}
\eer
Therefore, incorporating Eq.(\ref{msq}) and Eq.(\ref{psq}) in Eq.(\ref{R1}) we finally have :
\beq
R_{\mu\nu} = R_{\mu_\perp \nu} + R_{\mu_\parallel \nu} \,. 
\eeq
In writing this we have defined,
\ber
R_{\mu_\perp \nu} 
&=& g_{\mu_\perp \nu} g_{\twidle{\alpha_\parallel} \beta_\parallel} \,
    (p^{\alpha_\parallel} \, q^{\beta_\parallel} \, \tan  e{\cal B}s
    + q^{\alpha_\parallel} \, p^{\beta_\parallel} \, \tan  e{\cal B}s^{\prime}) \nonumber \\
&-& g_{\mu \beta_\perp} \, g_{\nu \twidle{\alpha_\parallel}} \, 
    q^{\beta_\perp} \, p^{\alpha_\parallel} \,
    (\tan e{\cal B}s - \tan e{\cal B}s^{\prime}) \nonumber \\ 
&+& g_{\mu \beta_\perp} \, g_{\nu \twidle{\alpha_\parallel}} \, 
    q^{\beta_\perp} \, q^{\alpha_\parallel} \,
    \frac{\tan^2  e{\cal B}s^{\prime} \, \sec^2 e{\cal B}s} 
    {\tan  e{\cal B}s + \tan  e{\cal B}s^{\prime}} \,.
\label{R_odd_perp}
\eer
It is evident that $R_{\mu_\parallel \nu} \, q^\nu = 0$ in accordance with Eq.(\ref{gi_pi5}).
On the other hand, we have,
\beq
R_{\mu_\parallel \nu} = R^{\rm A}_{\mu_\parallel \nu} + R^{\rm B}_{\mu_\parallel \nu} \,, 
\eeq
such that,
\ber
R^A_{\mu_\parallel \nu} 
&=& \varepsilon_{\mu \nu 1 2} \, (\tan  e{\cal B}s + \tan e{\cal B}s^{\prime}) 
    p^2_\parallel \nonumber \\
&+& (g_{\mu \twidle{\alpha_\parallel}} \, g_{\nu \beta_\parallel} 
    + g_{\mu \beta_\parallel} \, g_{\nu \twidle{\alpha_\parallel}} ) \,
    p^{\alpha_\parallel} \, p^{\beta_\parallel} 
    \, (\tan  e{\cal B}s + \tan  e{\cal B}s^{\prime}) \nonumber \\
&+& g_{\mu \twidle{\alpha_\parallel}} \, g_{\nu \beta_\parallel} \,
    p^{\alpha_\parallel} \, q^{\beta_\parallel} \, \tan  e{\cal B}s
    + (g_{\mu \twidle{\alpha_\parallel}} \, g_{\nu \beta_\parallel} 
    - g_{\mu_\parallel \nu} g_{\twidle{\alpha_\parallel} \beta_\parallel}) \,
    q^{\alpha_\parallel} \, p^{\beta_\parallel} \, \tan  e{\cal B}s^{\prime} \nonumber \\
&-& g_{\mu \twidle{\alpha_\parallel}} \, g_{\nu \beta_\perp} \,
    q^{\beta_\perp} \, p^{\alpha_\parallel} 
    (\tan e{\cal B}s - \tan e{\cal B}s^{\prime}) \,,
\eer
and
\ber
R^B_{\mu_\parallel \nu}
&=& \left(g_{\mu \beta_\parallel} \, g_{\nu \twidle{\alpha_\parallel}}  
    - g_{\mu_\parallel \nu} g_{\twidle{\alpha_\parallel} \beta_\parallel}\right) \,
    p^{\alpha_\parallel} \, q^{\beta_\parallel} \, \tan  e{\cal B}s 
    + g_{\mu \beta_\parallel} \, g_{\nu \twidle{\alpha_\parallel}} \,
    q^{\alpha_\parallel} \, p^{\beta_\parallel} \, \tan  e{\cal B}s^{\prime} \nonumber \\
&+& g_{\mu \twidle{\alpha_\parallel}} \, g_{\nu \beta_\perp} \,
    q^{\beta_\perp} \, q^{\alpha_\parallel} \,
    \frac{\tan^2  e{\cal B}s^{\prime} \, \sec^2 e{\cal B}s} 
    {\tan  e{\cal B}s + \tan  e{\cal B}s^{\prime}} 
    - \varepsilon_{\mu \nu 1 2} \, 
    \frac{\sec^2 e{\cal B}s \, \tan^2 e{\cal B}s^{\prime}}
    {\tan e{\cal B}s + \tan  e{\cal B}s^{\prime}} q^2_\perp \,.
\eer
Again, it is obvious that $R^{\rm B}_{\mu_\parallel \nu} q^\nu = 0$. In its present form, 
$R^B_{\mu_\parallel \nu}$ does not render itself to obvious gauge-invariance. However, 
the theory dictates that $R^{\rm A}_{\mu_\parallel \nu} q^\nu$ should vanish. And we have 
shown in the appendix~(\ref{app:gauge}) that it is indeed so. Therefore, the complete 
gauge-invariant expression for $R_{\mu \nu}$ is given by,
\ber
R_{\mu\nu}
&=& \varepsilon_{\mu \nu 1 2} \, (\tan  e{\cal B}s + \tan e{\cal B}s^{\prime}) 
    p^2_\parallel \nonumber \\
&+& \left(\varepsilon_{\mu \alpha_\parallel 12} \; g_{\nu \beta_\parallel} 
    + g_{\mu \beta_\parallel} \; \varepsilon_{\nu \alpha_\parallel 12}\right ) \,
    p^{\alpha_\parallel} \, p^{\beta_\parallel}
    \, (\tan  e{\cal B}s + \tan  e{\cal B}s^{\prime}) \nonumber \\
&+& \left(\varepsilon_{\mu \alpha_\parallel 12} \; g_{\nu \beta_\parallel} 
    + g_{\mu \beta_\parallel} \; \varepsilon_{\nu \alpha_\parallel 12}  
    - g_{\mu \nu} \; \varepsilon_{\alpha_\parallel \beta_\parallel 12}\right) \nonumber \\
&&  \times \left(p^{\alpha_\parallel} \, q^{\beta_\parallel} \, \tan  e{\cal B}s
    + q^{\alpha_\parallel} \, p^{\beta_\parallel} \, \tan  e{\cal B}s^{\prime}\right) \nonumber \\
&-& \left(\varepsilon_{\mu \alpha_\parallel 12} \; g_{\nu \beta_\perp} \,
    + g_{\mu \beta_\perp} \; \varepsilon_{\nu \alpha_\parallel 12}\right) \, 
    q^{\beta_\perp} \, p^{\alpha_\parallel} 
    \, (\tan e{\cal B}s - \tan e{\cal B}s^{\prime}) \nonumber \\ 
&+& \left(\varepsilon_{\mu \alpha_\parallel 12} \; g_{\nu \beta_\perp} \,
    + g_{\mu \beta_\perp} \; \varepsilon_{\nu \alpha_\parallel 12}\right) \, 
    q^{\beta_\perp} \, q^{\alpha_\parallel} 
    \, \frac{\tan^2  e{\cal B}s^{\prime} \, \sec^2 e{\cal B}s} 
    {\tan  e{\cal B}s + \tan  e{\cal B}s^{\prime}} \nonumber \\
&-& \varepsilon_{\mu \nu 1 2} \, 
    \frac{\sec^2 e{\cal B}s \, \tan^2 e{\cal B}s^{\prime}}
    {\tan e{\cal B}s + \tan  e{\cal B}s^{\prime}} q^2_\perp \,,
\label{R2}
\eer
where we have used the identity 
\ber
g_{\mu \twidle{\alpha_\parallel}} \, a^{\alpha_\parallel} 
= \varepsilon_{\mu \alpha_\parallel 12} \, a^{\alpha_\parallel} \,, 
\eer
valid for any vector $a^\alpha$. 
\section{The Weak Field Limit}{\label{weak}}
Retaining terms up-to $O({\cal B})$ in Eq.(\ref{R2}) we have :
\ber
\Pi^5_{\mu\nu}(q) (O({\cal B}))
&=& - 4e^3 {\cal B} \int \frac{d^4p}{(2\pi)^4} \; X(\beta, q, p)
      \int_{-\infty}^\infty ds\; e^{\Phi(p,s)} \,
      \int_{-\infty}^\infty ds^{\prime}\; e^{\Phi(p^{\prime},s^{\prime})} \nonumber \\
&\times&  [\, \varepsilon_{\mu \nu 1 2} \, (s + s') p^2_\parallel 
    + \left(\varepsilon_{\mu \alpha_\parallel 12} \; g_{\nu \beta_\parallel} 
    + g_{\mu \beta_\parallel} \; \varepsilon_{\nu \alpha_\parallel 12}\right ) \,
    p^{\alpha_\parallel} \, p^{\beta_\parallel} (s+s') \nonumber \\
&+& \left(\varepsilon_{\mu \alpha_\parallel 12} \; g_{\nu \beta_\parallel} 
    + g_{\mu \beta_\parallel} \; \varepsilon_{\nu \alpha_\parallel 12}  
    - g_{\mu \nu} \; \varepsilon_{\alpha_\parallel \beta_\parallel 12}\right) 
      \left(p^{\alpha_\parallel} \, q^{\beta_\parallel} \, s
    + q^{\alpha_\parallel} \, p^{\beta_\parallel} \, s' \right) \nonumber \\
&-& \left(\varepsilon_{\mu \alpha_\parallel 12} \; g_{\nu \beta_\perp} \,
    + g_{\mu \beta_\perp} \; \varepsilon_{\nu \alpha_\parallel 12}\right) \, 
    q^{\beta_\perp} \, p^{\alpha_\parallel} \, (s-s')\nonumber \\
&+& \left(\varepsilon_{\mu \alpha_\parallel 12} \; g_{\nu \beta_\perp} \,
    + g_{\mu \beta_\perp} \; \varepsilon_{\nu \alpha_\parallel 12}\right) \, 
    q^{\beta_\perp} \, q^{\alpha_\parallel} \, \frac{{s'}^2}{s+s'} 
    - \varepsilon_{\mu \nu 1 2} \, q^2_\perp \, \frac{{s'}^2}{s+s'}] \,.
\label{weak_field}
\eer
This entire expression {\it vanishes} upon integration, as has been shown in 
appendix~\ref{app:int}. Therefore, to $O({\cal B})$ the electro-magnetic vertex 
function is simply:
\beq
\Gamma_{\nu} = - \frac{1}{\sqrt{2}e} G_F \gamma^{\mu} (1 - \gamma_5) \,g_{\rm V} \Pi_{\mu \nu}(O({\cal B})) \,,
\eeq
where, $\Pi_{\mu \nu}(O({\cal B}))$ is given by :
\ber
\Pi_{\mu\nu}(O({\cal B})) 
&=& - \, 4ie^3 {\cal B} \varepsilon_{\mu\nu\alpha_\parallel\beta} q^\beta \, 
    \int \frac{d^4p}{(2\pi)^4} \, X(\beta, q, p) 
    \int_{-\infty}^\infty ds \; e^{\Phi(p,s)} \nonumber\\
&&  \times \, \int_{-\infty}^\infty ds' \; e^{\Phi(p',s')}
    \times \, p^{\widetilde\alpha_\parallel} \, \left( s + s' -  \frac{2 s s'}{s+s'} \right) \,.
\eer
For a detailed discussion on the properties of $\Pi_{\mu \nu}(O({\cal B}))$ in various 
background media see \cite{qft2}. \\

\ni
In this context, it would be worthwhile to compare our results with the case of a 
non-magnetic thermal plasma. Following the formulation in section~\ref{form} we find 
that the absorptive part of the 1-loop polarization tensor (the axial-vector--vector 
interaction) for a non-magnetic thermal plasma is given by Eq.(\ref{Pi5}) where,
\ber
iS_{\rm im} = (1/2 - \eta_F(p)) \int_{-\infty}^\infty ds\; e^{\Phi_0(p,s)} C_0(p,s) \,.
\eer
with,
\ber
\Phi_0(p,s) &=& is \left( p^2 - m^2 \right) - \epsilon |s| \,, \\
C_0(p,s) &=& \rlap/p + m \,.
\eer
With the above definitions, we find that,
\ber
\Pi^5_{\mu\nu}(q, {\cal B}=0) 
&& = - 16i \pi^2 \, e^2 \, \varepsilon_{\mu \nu \alpha \beta} \, q^\beta \nonumber \\
&&   \times \, \int \frac{d^4p}{(2\pi)^4} \, p^\alpha \, X(\beta, q, p) 
     \, \delta(p^2 - m^2) \, \delta({p'}^2 - m^2) \,.
\label{Pi5_thermal}
\eer
Therefore, the absorptive part of the polarization tensor, for a non-magnetic thermal 
plasma, is given by Eq.(\ref{Pi5_thermal}). Evidently, the same is true for a weakly 
magnetized plasma, to linear order in the strength of the magnetic field. 
\section{Conclusion}{\label{concl}}
In this work, we have considered massless neutrinos. However, recent observations 
indicate that the neutrinos may have mass. Nevertheless, our present treatment can be 
modified for massive neutrinos following the method adopted in \cite{nie3}.  \\
 
\ni
It is important to note that the correction to the absorptive part of the 
axial-vector--vector amplitude due to the presence of a magnetic field is zero to the 
linear order in the field strength compared to the case of a non-magnetic thermal plasma. 
Therefore, it indicates that the absorptive part of the axial-vector--vector amplitude 
is not enhanced, to the first order, by the presence of a magnetic field contrary to a 
naive expectation. It should be emphasized that this result is valid for the case of a 
weak magnetic field. However, as has been mentioned before, the magnetic field in most 
astrophysical systems are well below the critical field value. Therefore, our result 
would be pertinent to such situations.  \\

\ni
It also needs to be emphasized that unlike in the case of the real part (which gives 
the effective charge of the neutrinos) where a magnetic field breaks the isotropy of 
space there is no such introduction of a preferential direction in the case of 
absorption. Therefore, even though the effective charge of the neutrinos picks out 
the direction of the external magnetic field, to $O({\cal B})$ the absorption processes 
do not see any direction dependence.
\section*{Acknowledgments}

We would like to thank Palash B. Pal for helpful discussions and Kaushik Bhattacharya 
for pointing out some mistakes in the manuscript. In addition, we thank IUCAA, Pune, 
where SK was a post-doctoral fellow and IIT, Kanpur, where SD was an undergraduate 
student at the time of preparation of this manuscript.
\appendix
\section{Proof of gauge invariance}{\label{app:gauge}}
In order to show the gauge invariance of $\Pi^5{\mu_\parallel \nu}$ let us consider 
the case of $\mu_\parallel = 3$, first. Then we shall have,
\ber
{\Pi^5_{3 \nu}(q)}^{\rm A} \, q^\nu
&=& - 4e^2 \int \frac{d^4p}{(2\pi)^4} \; X(\beta, q, p) 
   \int_{-\infty}^\infty ds\; e^{\Phi(p,s)} \,
   \int_{-\infty}^\infty ds^{\prime}\; e^{\Phi(p^{\prime},s^{\prime})} \nonumber \\
&& [\, p^0 \, (q^2_\parallel + 2 p.q) \, (\tan e{\cal B}s + \tan e{\cal B}s^{\prime})
    - p^0 \, q^2_\perp \, (\tan e{\cal B}s - \tan e{\cal B}s^{\prime}) \,] \,.
\eer
Now, from the definition of $\Phi$, it follows that, apart from the small convergence factors, 
\ber
{i \over e{\cal B}} \, \left(\Phi(p,s) + \Phi(p',s')\right)
&=& \left( p_\parallel^{\prime2} + p_\parallel^2 - 2m^2 \right) \xi 
   - \left(p_\parallel^{\prime2} - p_\parallel^2 \right) \zeta \nonumber \\
&& - p_\perp^{\prime2} \tan (\xi-\zeta) - p_\perp^2 \tan (\xi+\zeta) \,,
\eer
where we have defined the parameters
\ber
\xi &=& \frac12 e{\cal B}(s+s') \,, \nonumber\\*
\zeta &=& \frac12 e{\cal B}(s-s') \,.
\label{xizeta}
\eer
Thus,
\ber
\left(p_\parallel^{\prime2} - p_\parallel^2 \right) e^{\Phi(p,s) + \Phi(p',s')} 
&=& \left( ie{\cal B} {d\over d\zeta} + p_\perp^{\prime2} \sec^2 (\xi-\zeta) 
           - p_\perp^2 \sec^2 (\xi+\zeta) \right) \nonumber \\ 
&&  \times \, e^{\Phi(p,s) + \Phi(p',s')} \,.
\label{C1par}
\eer
It should be noted that, $q^2_\parallel + 2 q.p_\parallel = {p'}^2_\parallel - p^2_\parallel$.
Hence,
\ber
{\Pi^5_{3 \nu}(q)}^{\rm A} \, q^\nu 
&=& - 4e^2 \int \frac{d^4p}{(2\pi)^4} \; X(\beta , k, p) \, p^0 
    \int_{-\infty}^\infty ds\; \int_{-\infty}^\infty ds^{\prime}\nonumber \\ 
&&  \times \, \large\{ \, (\tan e{\cal B}s + \tan e{\cal B}s^{\prime}) 
   \left( ie{\cal B} {d\over d\zeta} 
    + p_\perp^{\prime2} \sec^2 (\xi-\zeta) - p_\perp^2 \sec^2 (\xi+\zeta) \right)  \nonumber \\
&& - k^2_\perp \, (\tan e{\cal B}s - \tan e{\cal B}s^{\prime}) \,\large\} 
    e^{\Phi(p,s)} \, e^{\Phi(p^{\prime},s^{\prime})} \,.
\eer
Using Eq.(\ref{psq}) and Eq.(\ref{p'sq}) this can be further modified to :
\ber
{\Pi^5_{3 \nu}(q)}^{\rm A} \, q^\nu 
&=& - \, 4e^2 \, \int \frac{d^4p}{(2\pi)^4} \; X(\beta , k, p) \, p^0 
    \int_{-\infty}^\infty ds\; \int_{-\infty}^\infty ds^{\prime}\nonumber \\
&& (\tan e{\cal B}s + \tan e{\cal B}s^{\prime}) 
    ({d\over d\zeta} + \sec^2 e{\cal B}s - \sec^2 e{\cal B}s') \ 
    e^{\Phi(p,s)} \, e^{\Phi(p^{\prime},s^{\prime})} \nonumber \\
&& - \, 4e^2 q^2_\perp \, \int \frac{d^4p}{(2\pi)^4} \; X(\beta, k, p) \, p^0 
   \int_{-\infty}^\infty ds\; e^{\Phi(p,s)} \,
    \int_{-\infty}^\infty ds^{\prime}\; e^{\Phi(p^{\prime},s^{\prime})} \nonumber \\
&& [ \, \frac{\sec^2 e{\cal B}s' \, \tan^2 e{\cal B}s}{\tan e{\cal B}s + \tan e{\cal B}s'} 
   - \frac{\sec^2 e{\cal B}s \, \tan^2 e{\cal B}s'}{\tan e{\cal B}s + \tan e{\cal B}s'} 
   - (\tan e{\cal B}s - \tan e{\cal B}s^{\prime}) \,] \nonumber \\
\eer
which vanishes identically. Hence, ${\Pi^5_{3 \nu}(q)}^{\rm A}$ satisfies 
Eq.(\ref{gi_pi5}). The gauge invariance for ${\Pi^5_{0 \nu}(q)}^{\rm A}$ can be shown 
in a similar fashion. 
\section{Evaluation of the integrals}{\label{app:int}}
From the definition of $\Phi$ it follows that,
\ber
\Phi(p,s) +  \Phi(p',s') 
= it\,(p^2 + p.q + q^2/2 - m^2-i \epsilon_1)
 + it'\,( p.q + q^2/2 -i\epsilon_2) \,,
\eer
where $t=s+s'$ and $t'=s'-s$. Using this Eq.(\ref{weak_field}) can be rewritten in the 
following form,
\ber
\Pi^5_{\mu\nu}(q)
&=& - 4e^3 {\cal B} \int \frac{d^4p}{(2\pi)^4} \; X(\beta, q, p) \nonumber \\
&\times& \int_{-\infty}^\infty dt \int_{-\infty}^\infty dt' 
          \, ( A \, t + B \, t' + C \, \frac{{t'}^2}{t}) 
          \, e^{\Phi(p,s)} \, e^{\Phi(p',s')} \,.
\label{pi5_app}
\eer
where $A, B, C$ are functions of $p$ and $q$. Now,
\ber
&& \int_{-\infty}^\infty dt \, t \, e^{it( p^2 + p.q + q^2/2 - m^2 -i\epsilon_1)} \nonumber \\
&& = 4 \delta(p^2 + p.q + q^2/2 - m^2) \, (p^2 + p.q + q^2/2 - m^2) \nonumber \\
&&   \; \; \; \; ((p^2 + p.q + q^2/2 - m^2)^2 + \epsilon_1^2)^{-1} \,
\eer
where we have used the identity:
\beq
\frac{1}{a \pm i \epsilon} = {\cal P}(a) \mp i \pi \delta(a) \,, 
\eeq
$\cal P$ being the principal value and the integral:
\beq
\int^\infty_{-\infty} dx \, x \, e^{i(a - ib)x} = \, \frac{4 i ab}{(a^2 + b^2)^2} \,,
\label{eqnform}
\eeq
for $Re (b) > |Im (a)|$ ~\cite{grad}. Therefore, 
\ber
&& \int_{-\infty}^\infty dt \int_{-\infty}^\infty dt' \, t \, e^{\Phi(p,s)} \, e^{\Phi(p',s')} \nonumber \\
&& = \delta(p^2 + p.q + q^2/2 - m^2) 
    \,\frac{2(p^2 + p.q + q^2/2 - m^2)}{(p^2 + p.q + q^2/2 - m^2)^2+(\epsilon_1)^2} \nonumber \\
&&  \times \, \int_{-\infty}^\infty dt' \, e^{it'(p.q + q^2/2 - i \epsilon_2)}
\label{B2}
\eer
Since the numerator and the argument of the delta function are the same, this integral 
vanishes upon $p$-integration, provided we take the limit $\epsilon_1 \rightarrow 0^+$ 
later.  It could be similarly argued that the second term in Eqn.(\ref{pi5_app}) vanishes 
upon $p$-integration. In case of the third term, an integration by parts for the $t'$-integral 
renders it to the form of Eq.(\ref{eqnform}) and the above argument then can be followed 
through to show that this also vanishes upon $p$-integration.\\
\beb

\bibitem{raff}
G.~G. Raffelt, {\it Stars as Laboratories for Fundamental Physics}
  (University of Chicago Press, 1996).

\bibitem{hari}
L.~L. DeRaad, K.~A. Milton and N.~D. Hari-Dass, {\it Phys. Rev.} {\bf D14}, 3326 (1976).

\bibitem{oliv}
J.~C. Olivo, J.~F. Nieves, and P.~B. Pal, {\it Phys. Rev.} {\bf D40}, 3679 (1989).

\bibitem{ioan}
A.~N. Ioannisian and G.~G. Raffelt, {\it Phys. Rev.} {\bf D55}, 7038 (1997).

\bibitem{hard}
S.~J. Hardy and D.~B. Melrose, {\it Publ. Astron. Soc. Aus.} {\bf 13}, 144 (1996).

\bibitem{orae}
V.~N. Oraevsky, V.~B. Semikoz and Y.~A. Smorodinsky, {\it JETP Lett.} {\bf 43}, 709 (1986).

\bibitem{shai}
R. Shaisultanov, {\it Phys. Rev.} {\bf D62}, 113005 (2000).

\bibitem{gsh1}
H. Gies and R. Shaisultanov, {\it Phys. Rev.} {\bf D63}, 73003 (2000).

\bibitem{gsh2}
H. Gies and R. Shaisultanov, {\it Phys. Lett.} {\bf B480}, 129 (2000).

\bibitem{shame}
R. Shaisultanov, {\it Phys. Rev. Lett.} {\bf 80}, 1586 (1998).

\bibitem{nie3}
J.~F. Nieves and P.~B. Pal, {\it Phys. Rev.} {\bf D49}, 1398 (1994).

\bibitem{alth}
T. Altherr and P. Salati, {\it Nucl. Phys.} {\bf B421}, 662 (1994).

\bibitem{qft3}
K. Bhattacharya, A.~K. Ganguly and S. Konar, {\it Phys. Rev.} {\bf D65}, 013007 (2002).

\bibitem{qft1}
A.~K. Ganguly, S. Konar and P.~B. Pal, {\it Phys. Rev.} {\bf D60}, 105014 (1999).

\bibitem{sahu}
J.~C. D'Olivo, J.~F. Nieves and S. Sahu, {\it Phys. Rev.} {\bf D67}, 025018 (2003).

\bibitem{schu}
C. Schubert, {\it Nucl. Phys.} {\bf B585}, 429 (2000).

\bibitem{qft2}
A.~K. Ganguly and S. Konar, {\it Phys. Rev.} {\bf D63}, 065001 (2001).

\bibitem{adas}
A. Das, {\it Finite Temperature Field Theory} (World Scientific, 1997).

\bibitem{kob1}
R. Kobes, {\it Phys. Rev.} {\bf D42}, 562 (1990).

\bibitem{nie1}
J.~F. Nieves, {\it Phys. Rev.} {\bf D42}, 4123 (1990).

\bibitem{land}
N.~P. Landsman and C.~G. Weert, {\it Phys. Rep.} {\bf 145}, 141 (1987).

\bibitem{schw}
J. Schwinger, {\it Phys. Rev.} {\bf 82}, 664 (1951).

\bibitem{tsai}
W.~Y. Tsai, {\it Phys. Rev.} {\bf D10}, 1342, 2699 (1974).

\bibitem{ditt}
W. Dittrich, {\it Phys. Rev.} {\bf D19}, 2385 (1979).

\bibitem{elmf}
P. Elmfors, D. Grasso and G. Raffelt, {\it Nucl. Phys.} {\bf B479}, 3 (1996).

\bibitem{grad}
I.~S. Gradshteyn and I.~M. Ryzhik, {\it Table of Integrals, Series and Products} 
(Academic Press, 2000).
\eeb

\end{document}